\documentclass[preprint]{aastex}
\usepackage{psfig}
\usepackage{epsfig}
\def\approxgt{\ifmmode \rlap{$>$}{}_{{}_{{}_{\textstyle\sim}}} \else%
$\rlap{$>$}{}_{{}_{{}_{\textstyle\sim}}}$\fi} 
\def\approxlt{\ifmmode \rlap{$<$}{}_{{}_{{}_{\textstyle\sim}}} \else%
$\rlap{$<$}{}_{{}_{{}_{\textstyle\sim}}}$\fi}
\begin{document}
\title{Discovery of an X--ray pulsar in the low--mass X-ray binary
2A~1822--371}

\author{Peter G. Jonker\altaffilmark{1}, Michiel van der Klis\altaffilmark{1}}

\altaffiltext{1}{Astronomical Institute ``Anton Pannekoek'',
University of Amsterdam, and Center for High-Energy Astrophysics,
Kruislaan 403, 1098 SJ Amsterdam; peterj@astro.uva.nl,
michiel@astro.uva.nl} 

\begin{abstract}
\noindent
We report the discovery of 0.59~s X--ray pulsations from the low--mass
X--ray binary, 5.57 hr dipping and eclipsing ADC source
2A~1822--371. Pulse arrival time analysis indicates a circular orbit
with {\it e} $<$0.03 (95\% confidence) and an asin{\it i} for the
neutron star of 1.006(5) lightseconds, implying a mass function of
$(2.03\pm0.03)\times10^{-2} M_\odot$.  The barycentric pulse period
was 0.59325(2)~s in 1996.270 and 0.59308615(5)~s in 1998.205,
indicating an average spin up with $\dot P/P =
(-1.52\pm0.02)\times10^{-4} {\rm yr}^{-1}$. For a magnetic field
strength of $\sim$1--5$\times10^{12}$ G as derived from the X--ray
spectrum the implied intrinsic X--ray luminosity is
$\sim$2--4$\times10^{37}{\rm erg\,s^{-1}}$.  The pulse amplitude is low,
but increases steeply as a function of energy from a sinusoidal
amplitude of 0.25\% in 2--5.4 keV to $\sim$3\% above 20 keV. We
discuss the constraints on the masses of the companion star and the
fact that several aspects of the energy spectrum are in qualitative
accordance with that of a strongly magnetised neutron star.

\end{abstract}


\section{Introduction}
\label{intro}
\noindent
The lightcurve of the low--mass X--ray binary (LMXB) 2A~1822--371
shows clear signs of orbital modulation in both the X--ray and optical
bands (\citealt{1981ApJ...247..994W}; \citealt{1979IAUC.3406....1S};
\citealt{1980ApJ...242L.109M}), with a period of 5.57
hours. \citet{1981ApJ...247..994W} showed that the X--rays are
emitted from an extended source, a so called Accretion Disk Corona
(ADC) which is periodically partially eclipsed by the companion star
(at orbital phase 0.0) as well as partially obscured by structure in
the accretion disk whose height above the orbital plane varies but is
greatest at phase 0.8 and least at phase 0.2
\citep{1981ApJ...247..994W}. The implied inclination is {\it i}
$>$70$^\circ$ \citep{1980ApJ...242L.109M}. The short orbital period
makes 2A~1822--371 a compact LMXB. If powered by a Roche lobe filling
main--sequence star the companion mass is 0.62$M_\odot$, however, the
companion spectrum is inconsistent with that of a normal K--star
(\citealt{harlaftis97}). \par
\noindent
The orbital period has been measured from eclipse timing to gradually
increase \citep{1990MNRAS.244P..39H}; the best ephemeris to date was
provided by \citet{parmaretal2001}. The observed X--ray spectrum is
complex and various models have been used to describe the data. With a
power law index of $\sim$1 \citep{parmaretal2001} the continuum is
harder than that of typical LMXBs which have power law indices of
1.5--2.5. There is also evidence for a strong soft component in the
1-10 keV range (e.g. \citealt{2001MNRAS.320..249H}; Parmar et
al. 2001). An upper limit on the presence of pulsations in the 1--30
keV band of 1\% was derived by \citet{1992MNRAS.258..457H}. \par
\noindent
Soon after the discovery of accreting X--ray pulsars
\citep{1971ApJ...167L..67G} it was realized that these are strongly
magnetized (B$>10^{12}$ G) neutron stars accreting matter from an
accretion disk (\citealt{1972A&A....21....1P};
\citealt{1973ApJ...184..271L}) or a stellar wind
\citep{1973ApJ...179..585D}. Whereas accretion--powered pulsars are
common in massive X--ray binaries, they are rare in LMXBs, a fact that
has been explained in terms of neutron star magnetic field decay
(presumably accretion--induced) in the generally much longer--lived
low--mass systems (\citealt{bhasri1995}). The lower field would allow
the disk to extend to close to the neutron star and spin it up to
millisecond periods. This is in accordance with binary evolutionary
models predicting that LMXBs are the progenitors of binary millisecond
radio pulsars (\citealt{radsri1982}; \citealt{1982Natur.300..728A};
see for a detailed description \citealt{bhatta1995}). This scenario
was confirmed by the discovery of the first accreting millisecond
pulsar, in the LMXB SAX~J1808.4--3658
(\citealt{1998Natur.394..344W}). In this Letter, we report the
discovery of pulsations in the LMXB 2A~1822--371, and describe our
measurements of both the orbital Doppler shifts and the spin--up of
the pulsar. We briefly discuss the constraints on the masses of the
two binary components and also the energy spectrum of the pulsar.

\section{Observations and analysis}
\label{analysis}
\noindent
We used 16 observations obtained on 1996 Sept. 26 and 27 (observations
1--5), 1998 June 28 and 29 (observations 6--11), and July 24 and 25
(observations 12--16) with the proportional counter array (PCA;
\citealt{jaswgi1996}) onboard the {\it Rossi X--ray Timing Explorer}
(RXTE) satellite \citep{brrosw1993}. The total amount of good data was
$\sim$73 ksec. All observations yielded data with a time resolution of
at least $2^{-13}$ s, in 64 energy bands covering the effective
2.0--60 keV energy range of RXTE.\par
\noindent
As part of a systematic search for pulsars in LMXBs (Jonker et al. in
prep.), a power spectrum of Solar System barycentered data was created
using an FFT technique. The Nyquist frequency was 64 Hz. A weak 0.59~s
pulsed signal was discovered first in the 2.0--60 keV power
spectrum. Investigation of the pulsed signal in various energy bands
and different sub--sets of the data showed that the signal--to--noise
ratio was highest in the 9.4--22.7 keV band of observations
12--16. Therefore, we initially used this energy band and subset of
the data for our analyses.  \par
\begin{figure*}
\centerline{\psfig{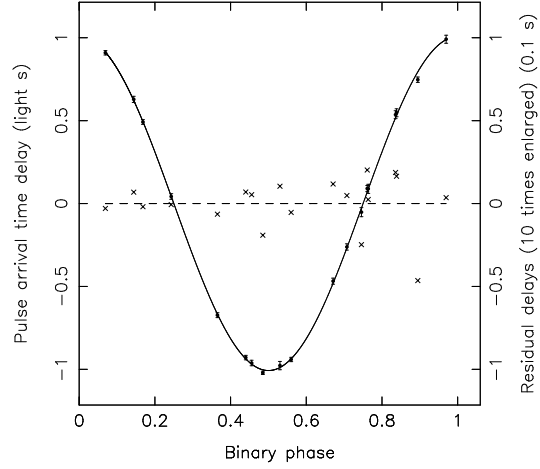}}\figcaption{The
arrival time delay in light seconds of the pulses due to the orbital
motion of the neutron star as a function of binary phase. Phase zero
is superior conjunction. Each dot represents $\sim$1500~s of data
obtained in observations 12--16. The sinusoid is the best fit to the
dots. The residuals of the fit (crosses) are shown at a 10 times
expanded scale. Error bars are shown for the dots; for clarity they
are omitted for the residuals. \label{delay}}
\end{figure*}

\noindent
We measured the Solar System barycentric pulse period in 19 data
segments of observations 12--16 each with a length of $\sim$1500~s
(half a typical RXTE contiguous data segment) using an epoch folding
technique. The period of the pulsar showed clear evidence of the 5.57
hour orbital modulation due to orbital Doppler shifts with an
amplitude corresponding to an asin{\it i} of about 1 light
second. Correcting for the orbital delays using the previously known
orbital ephemeris (Parmar et al. 2001) and our best measure of
asin{\it i} obtained from the pulse period analysis, we epoch--folded
each 1500~s segment in observations 12--16, and measured the phase of
each folded profile by fitting it with a sinusoid. The residual phases
were then fitted with a model using a constant pulse period and a
circular orbit. This satisfactorily described the observed
dependencies on both time and orbital phase. The best--fit orbital and
pulse parameters are given in Table~\ref{findings}.  The measured
pulse arrival times and the best--fit orbital delay curve are
displayed in Fig.~\ref{delay}.

\par
\noindent
Assuming our measured asin{\it i} and the orbital ephemeris of
\citet{parmaretal2001} we found for observations 1--5 a pulse period
of 0.59325(5)~s. This is significantly longer than that during
observations 12--16 (see Table~\ref{findings}), a conclusion that is
insensitive to the details of the orbital corrections.  From this
difference we derived a pulse period derivative of $\dot P =
(-2.85\pm0.04)\times 10^{-12} {\rm s\,s^{-1}}$. Due to the weakness of
the signal and the limited amounts of data we were not able to
phase--connect the data within observations 1--5 or 6--11, nor could
we maintain the pulse count between the epochs of observations 1--5,
6--11, or 12--16.  \par
\begin{figure*}
\centerline{\psfig{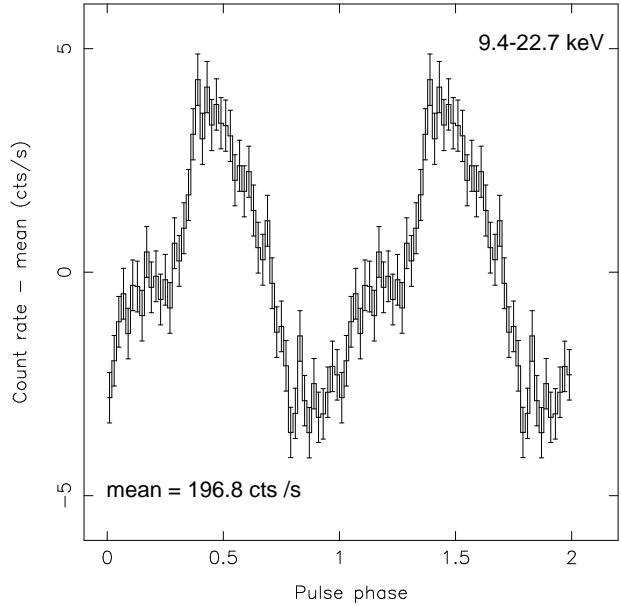}} \figcaption{The
measured pulse profile obtained from epoch folding the 9.4--22.7 keV
data of observations 12--16. The mean count rate (indicated) was
subtracted. For clarity two periods are plotted. The profile is
clearly non--sinusoidal. Phase zero is at HJD 2451019.4011752. The bin
size is $\sim$0.01~s. One sigma error bars are shown.
\label{shape}}
\end{figure*}
\begin{figure*}
\centerline{\psfig{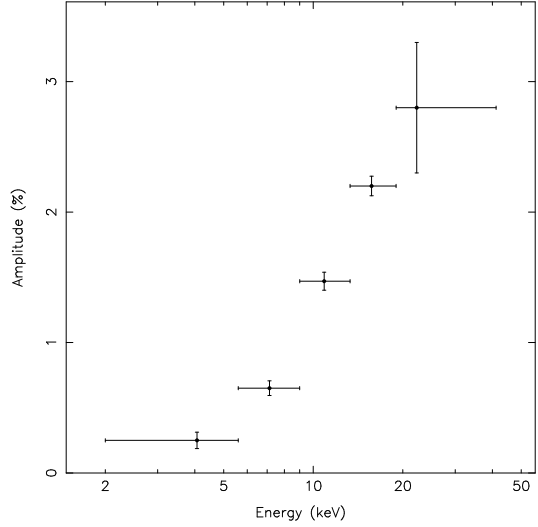}}
\figcaption{Pulse amplitude as a function of energy. The horizontal
bars denote the width of the energy bands, while the vertical bars
denote 1$\sigma$ uncertainties. The x--coordinate of the dots is the
weighted mean photon energy in each band. The pulse amplitude
increases steeply with energy.
\label{ampl_vs_energy}}
\end{figure*}

\noindent
Using the parameters in Table~\ref{findings} we folded 30 ksec of data
of observations 12--16 in the energy bands
2.0--5.4--9.4--13.8--22.7--60 keV to measure the pulse shape and the
pulse amplitude as a function of energy. The pulse profiles are
consistent with being the same in each energy band and did not change
significantly as a function of binary phase. The best pulse profile
was obtained combining the energy bands 9.4--13.8 keV and 13.8--22.7 keV
(see Fig.~\ref{shape}).  We fitted a single sinusoid to the profile in
each energy band to measure the amplitude. The pulse amplitude depends
strongly on energy, increasing from 0.25\%$\pm$0.06\% in the 2.0--5.4
keV band to 2.8\%$\pm$0.5\% in the 22.7--60 keV band
(Fig.~\ref{ampl_vs_energy}). The pulse amplitude was lower in
observations 1--5 than in observations 6--11 and 12--16 ($\sim$1.2\%
versus $\sim$1.7\% and $\sim$2.1\% in the 13.8--22.7 keV band,
respectively).  Although a single sinusoid is not a perfect
representation of the pulse profile this will not significantly affect
the derived pulse phase differences or pulse amplitude spectrum, as
the profile is constant within the errors.

\section{Discussion}
\begin{figure*}
\centerline{\psfig{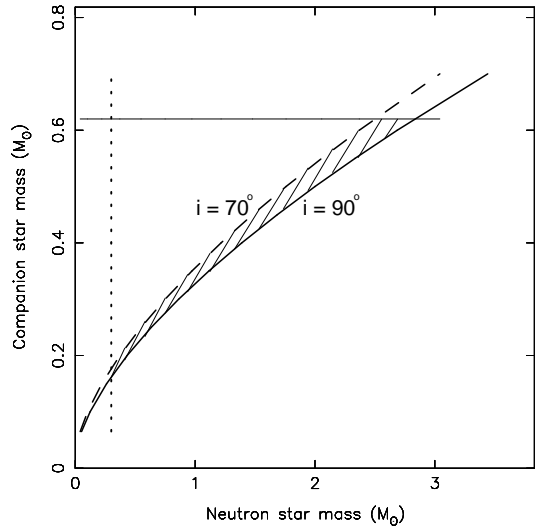}}
\figcaption{Companion star mass as a function of neutron star
mass. The system is located between the two curves representing {\it
i} $ = 70^\circ$ and {\it i} $ = 90^\circ$. The region to the left of
the dotted line is excluded (67\% confidence) due to the lower limit
on the radial velocity of the companion star
(\citealt{harlaftis97}). The horizontal line is the mass of the
companion assuming it is a Roche--lobe filling main sequence star. The
allowed region (shaded) assumes that the companion could be
undermassive, not more massive than this.
\label{mass}}
\end{figure*}

\noindent
Using data obtained with the RXTE satellite we have discovered 0.59~s
X--ray pulsations from the low--mass X--ray binary (LMXB) 2A~1822--371
with $\dot P/P = -1.5\times10^{-4} {\rm yr^{-1}}$. This is the sixth
LMXB to show pulsations (Table~\ref{comp}), the fourth whose orbital
pulse delay curve was measured, and after SAX~J1808.4--3658 only the
second compact LMXB (${\rm P_{orb} < 12\,hrs}$) for which this was
done. Contrary to SAX~J1808.4--3658, 2A~1822--371 is optically bright
and has a well--constrained inclination (because it is eclipsing),
which might allow for a future full binary solution. Before our
measurements, the nature of the compact object in 2A~1822--371 was
somewhat uncertain.  \citet{2001MNRAS.320..249H} showed that it could
either be a white dwarf, a neutron star or a low--mass black hole. Our
detection of pulsations, together with spin period changes on a
timescale of $\sim$10$^4$ years establishes that the compact object is
a neutron star. We derive a mass function for the companion star of
$(2.03\pm0.03)\times10^{-2} M_\odot$.  This, combined with the
knowledge of the inclination constrains the masses of the two
components to a small area in a plot of companion star mass versus
neutron star mass (the shaded region in Fig.~\ref{mass}).  If the
companion is a main sequence Roche--lobe filling star subject to the
usual lower main sequence mass--radius relation (\citealt{kipwei}) its
mass is 0.62$M_\odot$ (the horizontal line in Fig.~\ref{mass}). This
would imply a quite massive neutron star. Spectroscopic observations
provide a lower limit to the semi--amplitude of the radial velocity of
the companion star (\citealt{harlaftis97}). From that lower limit we
constrain the mass of the neutron star to be more than
$0.6_{-0.3}^{+1.0}M_\odot$ (the vertical line in Fig.~\ref{mass} at
0.3 $M_\odot$ represents the 67\% confidence limit). Furthermore, they
showed that the inner face of the companion star is 10~000--15~000 K
hotter than its back face. This is probably due to effects of X--ray
heating, which could render the companion significantly undermassive
for its size. For a 1.4 $M_\odot$ neutron star the companion has a
mass of 0.4--0.45 $M_\odot$ (see Fig.~\ref{mass}).
\par
\noindent
An estimate of the strength of the magnetic field depends among other
things on the source luminosity. The luminosity of 2A~1822--371 was
estimated (\citealt{1982ApJ...262..253M}) to be ${\rm L_X} \sim 1.1
\times$ $10^{35} {\rm erg\,s^{-1}}$ ${\rm (d/1kpc)^2}$. For a distance
of 2.5 kpc (\citealt{1982ApJ...262..253M}) this would lead to $\sim
10^{36} {\rm erg/s}$. However, since all observed X--rays are thought
to have been scattered by an ADC (see Section 1), the true source
luminosity may be as high as $\sim 10^{38} {\rm erg/s}$
(\citealt{1982ApJ...257..318W}). Such a high luminosity would be
consistent with the observed binary orbital period change
(\citealt{2001MNRAS.320..249H}). From the luminosity and the spin--up
rate the magnetic field can be determined
(cf. \citealt{1979ApJ...234..296G}); for ${\rm L_X \sim10^{38}}{\rm
erg\,s^{-1}}$ we derive a magnetic field strength, B, of
$\sim8\times10^{10}$ G, whereas for ${\rm L_X \sim10^{36}}{\rm
erg\,s^{-1}}$ B $\sim8\times10^{16}$ G, which implies that the
luminosity is probably not that low. If we assume that the neutron
star is spinning at its equilibrium period, then for ${\rm L_X
\sim10^{36}}{\rm erg\,s^{-1}}$ we find B$\sim5\times10^{10}$ G and for
${\rm L_X \sim10^{38}}{\rm erg\,s^{-1}}$ B$\sim5\times10^{11}$ G. In
all this we assumed $M_{ns} = 1.4 M_\odot$, $I=10^{45}$ g ${\rm
cm^2}$, and $R=10^6$ cm for the mass, moment of inertia, and radius of
the neutron star.  \par
\noindent
The X--ray spectrum of 2A~1822--371 was studied by various authors
(\citealt{1981ApJ...247..994W}; \citealt{1989MNRAS.239..715H};
\citealt{1992MNRAS.258..457H}; \citealt{2001MNRAS.320..249H};
\citealt{parmaretal2001}; \citealt{iariaetal2001}), using data
obtained with different satellites ({\it Einstein, EXOSAT, Ginga,
ASCA, RXTE}, and {\it BeppoSax}). \citet{parmaretal2001} discussed
several unusual features of the spectrum of 2A~1822--371 and although
Compton scattering in the ADC probably also affected the spectrum
(\citealt{1982ApJ...257..318W}), in principle some of these features
could be explained by the presence of a $\sim 10^{12}$ G pulsar
instead of 10$^{8}$--10$^{9}$ G neutron star. With a power law index
of $\sim$1 the continuum spectrum is much harder than that of LMXBs of
similar luminosity \citep{parmaretal2001}. This is, however, a common
power law index for X--ray pulsars (\citealt{1983ApJ...270..711W}).
The observed cut--off at $\sim$17 keV (\citealt{parmaretal2001}) could
also be explained by the presence of the pulsar. The cut--off energy
of pulsars is thought to be approximately half the cyclotron energy
(\citealt{maki1992}; see \citealt{whnapa1995} for a overview). The
strength of the B--field which can thus be derived from the cut--off,
assuming a redshift at the neutron star surface of 0.3, is
$\sim4\times10^{12}$ G.  The relation between the electron temperature
and the energy of the cyclotron resonance \citep{makishimaetal1999},
leads for a ${\rm kT_e}$ of $\sim$4--10 keV (\citealt{parmaretal2001};
\citealt{iariaetal2001}, although fit with a slightly different continuum
function than \citealt{makishimaetal1999}) to magnetic field estimates
of $\sim$ 1--5$\times10^{12}$ G, again assuming a redshift at the
neutron star surface of 0.3. These estimates of the magnetic field are
consistent with the estimates derived from the spin--up above. The
intrinsic source luminosity would be $\sim$2--4${\rm\times10^{37}
erg\,s^{-1}}$ given the $\dot P/P$ we measured.  \par
\noindent
The neutron star was found to spin up on a timescale of $\sim$6500
years; $\dot \nu$ is $(8.1\pm0.1)\times 10^{-12} {\rm
Hz\,s^{-1}}$. Comparing this $\dot \nu$ with that in the other LMXB
X--ray pulsars (Table~\ref{comp}) we note that the spin--up rate
measured over $\sim$666 days is large for an LMXB X--ray pulsar, but
that of the transient system GRO~J1744--28 is even larger
(\citealt{1996Natur.381..291F}). Recent observational evidence
summarized by \citet{1997ApJS..113..367B} reveals alternating episodes
of spin--up and spin--down in disk--fed neutron stars. If the
$\dot\nu$ we measured of 2A~1822--371 between 1996.270 and 1998.205
would be the average of multiple spin--up and spin--down episodes,
then the maximum spin--up rate would be even higher. However, episodes
of steady spin--up or spin--down lasting nearly a decade have been
observed in GX~1+4 and 4U~1626--67 (\citealt{1997ApJ...481L.101C};
\citealt{1997ApJ...474..414C}).  \par
\noindent
The increase in pulse amplitude with photon energy is steeper than has
been found for other low--mass X--ray pulsars (4U~1626--67,
\citealt{1977ApJ...217L..29R}; Her~X--1,
\citealt{1983ApJ...270..711W}; SAX~J1808.4--3658,
\citealt{1998Natur.394..344W}). Furthermore, the pulse amplitude is
low compared with other LMXB X--ray pulsars. Previous studies revealed
that in 2A~1822--371 scattering is important
(\citealt{1981ApJ...247..994W}; \citealt{1989MNRAS.239..715H};
\citealt{1992MNRAS.258..457H}; \citealt{parmaretal2001};
\citealt{2001MNRAS.320..249H}).  Multiple scatterings of the pulsed
emission in an ADC of 0.3$R_\odot$ (\citealt{1981ApJ...247..994W})
would have washed out the pulse due to light travel time
delays. Therefore, at least a portion of the ADC should not be very
optically thick. Compton scattering in such an ADC could explain the
observed pulse amplitude spectrum.

\acknowledgments \noindent This work was supported in part by the
Netherlands Organization for Scientific Research (NWO) grant
614-51-002. This research has made use of data obtained through the
High Energy Astrophysics Science Archive Research Center Online
Service, provided by the NASA/Goddard Space Flight Center. This work
was supported by NWO Spinoza grant 08-0 to E.P.J.van den Heuvel. PGJ
would like to thank Rob Fender and Jeroen Homan for carefully reading
an earlier version of the manuscript and Tiziana Di Salvo for
discussions on the X--ray spectrum.

\begin{deluxetable}{lr}
\tablecaption{Orbital parameters of 2A~1822--371. The number in
brackets indicates the 1$\sigma$ uncertainty in the last
digit. \label{findings}}

\tabletypesize{\normalsize} 
\tablecolumns{2}
\tablewidth{0pc}

\tablehead{ \colhead{} & \colhead{}} 
\startdata 
Barycentric pulse period (s) at 1998.205, $P_{1998}$ & 0.59308615(5)\\ 
Projected semimajor axis (light sec.), asin{\it i} & 1.006(5)\\ 
Orbital period (s), ${\rm P_{orb}}$ & 20054.240(6)$^a$\\ 
Epoch of superior conjunction (HJD), T & 2450993.27968(2)\\ 
Eccentricity, {\it e} (95\% confidence) & $<3.1 \times 10^{-2}$ \\
Mass function, $f_x(M_\odot)$ & $(2.03\pm0.03)\times10^{-2}$\\ 
\tableline
Barycentric pulse period (s) at 1996.270, $P_{1996}$ & 0.59325(2)\\ 
Pulse period derivative (s/s), $\dot P$ & $(-2.85\pm0.04) \times
10^{-12}$\\
\tablenotetext{a}{From orbital ephemeris of Parmar et al. (2001)}
\enddata
\end{deluxetable}

\begin{deluxetable}{llllll}
\tablecaption{Comparing the X--ray pulsars in LMXBs. \label{comp}}
\tabletypesize{\normalsize} 
\tablecolumns{6} 
\tablewidth{0pc}
\tablehead{\colhead{Source} & \colhead{$P_{Pulse}$ (s)} & \colhead{$\dot \nu$
(Hz/s)} & \colhead{$P_{orb}$ (days)} & \colhead{asin{\it i}} & \colhead{References}} 
\startdata 
SAX~J1808.4--3658 & 0.00249 & $<7\times10^{-13}$ & 0.0839 & 0.062809(1) & 1,2\\
GRO~J1744--28 & 0.467	& $1.2\times10^{-11}$
& 11.8 & 2.6324(1) & 3,4 \\
2A~1822--371 & 0.5931 & $8.1\times10^{-12}$ & 0.232
& 1.006(5) & 5 \\
Her~X--1 & 1.24 & $5\times10^{-13}$ & 1.70 & 13.1853(2) & 4,6 \\
4U~1626--67 & 7.66 & $8\times10^{-13}$ &
0.0289 & \nodata & 4,7\\
GX~1+4 & 120 & $6\times10^{-12}$ &
$\sim$304 & \nodata & 4,8,9,10\\

\tablenotetext{1}{\citealt{1998Natur.394..344W}}
\tablenotetext{2}{\citealt{1998Natur.394..346C}}
\tablenotetext{3}{\citealt{1996Natur.381..291F}}
\tablenotetext{4}{\citealt{1997ApJS..113..367B}}
\tablenotetext{5}{this paper}
\tablenotetext{6}{\citealt{1972ApJ...174L.143T}}
\tablenotetext{7}{\citealt{1977ApJ...217L..29R}}
\tablenotetext{8}{\citealt{1971ApJ...169L..17L}}
\tablenotetext{9}{\citealt{1997ApJ...481L.101C}}
\tablenotetext{10}{\citealt{1999ApJ...526L.105P}}

\enddata
\end{deluxetable}

\end{document}